\documentstyle[12pt]{article}
\begin{document}
\begin{titlepage}
\hspace{9cm} ULB--PMIF--93/11

\vspace{2.5cm}
\begin{centering}

{\huge Renormalization of gauge invariant operators and
anomalies in Yang-Mills theory}\\
\vspace{1cm}
{\large Glenn Barnich$^\dagger$ and Marc Henneaux$^*$\\
Facult\'e des Sciences, Universit\'e Libre de Bruxelles,\\
Campus Plaine C.P. 231, B-1050 Bruxelles, Belgium}\\
\end{centering}
\vspace{.25cm}
\begin{abstract}
A long-standing conjecture on the structure of renormalized, gauge invariant,
integrated operators of arbitrary  dimension in Yang-Mills theory is
established. The general solution of the consistency condition for
anomalies with sources included is also derived. This is achieved by computing
explicitely the cohomology of the full unrestricted BRST operator in
the space of local polynomial functionals with ghost number equal to zero or
one. The argument does not use power counting and is purely cohomological.
It relies crucially on standard properties of the antifield formalism.
\end{abstract}
\vspace{.25cm}
{\footnotesize ($^\dagger$)Aspirant au Fonds National de la Recherche
Scientifique (Belgium)\\
($^*$)Also at Centro de Estudios
Cient\'\i ficos de Santiago, Chile.}

\end{titlepage}
\def\carre{\vbox{\hrule\hbox{\vrule\kern 3pt
\vbox{\kern 3pt\kern 3pt}\kern 3pt\vrule}\hrule}}
\newtheorem{theorem}{\indent Theorem}
\newtheorem{corollary}{\indent Corollary}
\newtheorem{lemma}{\indent Lemma}

The antifield formalism \cite{Bat_Vil} is well known to be extremely
useful in the study of theories with a complicated off-shell gauge
structure like supergravity or string field theory. Much less appreciated
is the fact that the ideas underlying the antifield construction
-in particular the central feature that the antifields implement the
equations of motion in cohomology \cite{Fis_Hen_Sta_Tei}- are actually quite
helpful already in the analysis of the Yang-Mills field. An illustration of
this property has been given in \cite{Hen}, where the demonstration of
an old theorem by Joglekar and Lee \cite{Jog_Lee} on the renormalization of
local gauge invariant operators was streamlined using concepts from the
antifield formalism. In this letter, we extend the analysis of \cite{Hen}
by proving a long-standing conjecture originally due to Kluberg-Stern
and Zuber \cite{Klu_Zub,Voro,Zin} concerning the structure of renormalized,
{\it integrated}, gauge invariant operators (like $tr  F_{\mu\nu}^2$ at zero
momentum, i.e., $\int tr F_{\mu\nu}^2\ d^4x$).
We also provide the general solution of the anomaly equation
{\it with sources included}. Our approach does
not use power counting and is purely cohomological. Hence, it is valid
for operators of arbitrarily high physical dimension, for which our results
control both the possible counterterms and the anomalies.

Thanks to the effort of various people over many years, it has been shown
that most questions about the quantum properties of Yang-Mills models
can be reformulated as algebraic questions involving graded differential
algebras. It would be out of place to review here the huge body of work
that has gone into that problem. We shall rather refer to the  recent
historical survey given in reference \cite{Sto}.

As recalled there, the analysis of renormalized, integrated, gauge
invariant operators and of anomalies ultimately boils down to the
calculation of the cohomology of the BRST differential $s$ in the space
of integrated polynomials in the Yang-Mills potential $A^a_\mu$, the
ghosts $C^a$, the matter fields $y^i$ (belonging to some representation
of the gauge group), the antifields $A^{*\mu}_a$, $C^*_a$, and $y^*_i$
(also called ``sources for the BRST variations"), as well as their
spacetime derivatives. More precisely, the question is to find the general
solution of the equation
\begin{eqnarray}
sA=0,\label{fund_eq}
\end{eqnarray}
where $s$ is the BRST differential and where $A$ is the integral of a
polynomial in $A^a_\mu,C^a,y^i,
A^{*\mu}_a,C^*_a,y^*_i$ and their derivatives of ghost number zero or one,
\begin{eqnarray}
A=\int a\ d^4x,\ \ gh\ A=0\ or\ 1.
\end{eqnarray}
We shall allow for the presence of abelian factors in the gauge group but
we shall assume, however, that each abelian gauge field is coupled to at
least one charged matter field (no free abelian gauge field).
The following theorems are the central results reported in this letter.
\begin{theorem}
: the general solution of (\ref{fund_eq}) with ghost number zero is given by
\begin{eqnarray}
A=\int \bar a\ d^4x + s \int b\ d^4x \label{lagrangians}
\end{eqnarray}
where $\bar a$ is an invariant polynomial in the field strengths, the
matter fields, and their covariant derivatives. If the gauge group has
abelian factors, there are in addition extra solutions given by
\begin{eqnarray}
A=k_A^\Delta\int (j^\mu_\Delta A^A_\mu + t^a_{\Delta\mu}A^{*\mu}_a C^A
+ t^i_\Delta y_i^*C^A)\ d^4x \label{currents}
\end{eqnarray}
where: (i) $k^\Delta_A$ are arbitrary constants~; (ii) $A^A_\mu$ are the
abelian gauge fields and $C^A$ the abelian ghosts~;
and (iii) $j^\mu_\Delta$ are gauge invariant
conserved currents (for each $\Delta$),
\begin{eqnarray}
\partial_\mu j^\mu_\Delta = t^a_{\Delta\mu}{\delta {\cal L}\over\delta
A^a_\mu} + t^i_\Delta {\delta {\cal L}\over\delta y^i}
\end{eqnarray}
\end{theorem}
\begin{theorem}
: the general solution of (\ref{fund_eq}) with ghost number one is given by
\begin{eqnarray}
A=\int c_1\ d^4x + s \int b\ d^4x \label{anomalies}
\end{eqnarray}
with
\begin{eqnarray}
c_1=tr ( C\partial_\mu A_\nu\partial_\lambda A_\rho
+{3\over 2} C\partial_\mu A_\nu A_\lambda A_\rho)
\varepsilon^{\mu\nu\lambda\rho},
\label{Adler-Bardeen}
\end{eqnarray}
(Adler-Bardeen-Bell-Jackiw anomaly).
Again, if there are abelian factors,
there are further solutions and one must replace (\ref{anomalies}) by
\begin{eqnarray}
A=\int c_1\ d^4x +\int \mu_A C^A\ d^4x +\int c_2\ d^4x +\nonumber\\
\int c_3\ d^4x +s\int b\ d^4x
\label{DRM}
\end{eqnarray}
where (i) $\mu_A$ are invariant polynomials in the field strengths, the
matter fields and their covariant derivatives~; (ii) $c_2$ does not
depend on the antifields
\begin{eqnarray}
c_2 = k_A F^A_{\mu\nu} tr( C \partial_\lambda
A_\rho)\varepsilon^{\mu\nu\lambda\rho}\ ;
\label{gammaa_mod_d}
\end{eqnarray}
and (iii) $c_3$ is given by
\begin{eqnarray}
c_3= k_{AB}^\Delta (j^\mu_\Delta A^A_\mu C^B + {1\over 2}
t^a_{\Delta\mu}A^{*\mu}_a C^A C^B \nonumber\\
+ {1\over 2}t^i_\Delta y_i^*C^A C^B),\ \
k_{AB}^\Delta=-k_{BA}^\Delta \label{anom_suppl}.
\end{eqnarray}
As in (\ref{currents}), the $j^\mu_\Delta$ are conserved currents.
\end{theorem}

The solution (\ref{lagrangians}) is BRST-trivial if and only if
$\bar a$ vanishes on-shell (the antifields enable
one to rewrite a polynomial that vanishes on-shell as a
$s$-variation).  The solutions (\ref{currents})
and (\ref{anom_suppl}) are trivial if and only if the conserved
current $j^\mu _\Delta $ is trivial, i.e., equal on-shell to an
identically conserved total divergence.  The
conjecture of Kluberg-Stern and Zuber
mentioned above follows from a direct
inspection of the solutions given in theorems 1 and 2:
\begin{corollary}
: for a semi-simple gauge group, the general solution of
$s\int a\ d^4x =0$ with ghost number zero is, up to $s$-boundaries
(i.e., $s$-exact terms), equal to the integral of a gauge invariant
polynomial in the field strengths, the matter fields and their covariant
derivatives.
\end{corollary}
\begin{corollary}
: similarily, the only possible anomaly for
a semi-simple gauge group is of the
Adler-Bardeen-Bell-Jackiw type.
\end{corollary}
Thus, the conjecture holds for QCD or grand unified models. However,
if the gauge group has abelian factors, like in the standard model
with $U(1)\times SU(2) \times SU(3)$, then there are further
non-trivial solutions to $s\int a\ d^4x=0$ besides
(\ref{lagrangians}) and (\ref{anomalies}), (\ref{Adler-Bardeen}).
The anomaly solution (\ref{gammaa_mod_d}), which does not depend on
the antifields, has been known for some time. The other extra
solutions depend explicitly on the antifields and deserve some
comments.

(i) All the antifield dependent solutions are completely determined
(in cohomology) by their antifield independent component $A_0$,
which is a solution of $sA_0\approx 0$ where $\approx$ means equal
modulo the equations of motion). This is a standard result of
homological perturbation theory, which indicates how the antifield
dependent components follow recursively from the antifield
independent ones (see e.g. \cite{Hen_Tei}, chapter 8).
Thus, all the information about  (\ref{currents}) or (\ref{anom_suppl})
is respectively contained in $k_A^\Delta j^\mu_\Delta A^A_\mu$ or
$ k_{AB}^\Delta j^\mu_\Delta A^A_\mu C^B$.

(ii) The solutions (\ref{currents}) and (\ref{anom_suppl}) involve
explicitly conserved currents. An example of (\ref{anom_suppl}) has been
given in \cite{Bra}. While (\ref{anom_suppl}) needs at least
two abelian factors, the solution (\ref{currents}) exists already
with a single $U(1)$ provided there are non trivial conserved currents.
By the Noether theorem, a non trivial conserved
current exists for each non trivial rigid
symmetry (e.g., rigid $SU(N)$ flavor symmetry).
If all exact symmetries are gauge symmetries, however, all the
conserved currents are trivial and
the solutions (\ref{currents}) and (\ref{anom_suppl})
are also trivial.  If the spacetime symmetries are
not gauged, one may take for $j^\mu_\Delta$ the
energy momentum tensor $T^\mu_\nu$ (but the solution is then
non covariant).   It is not surprising that
the solutions of $sA=0$ contain information about the dynamics
through the appearance of conserved currents, since the role of the
antifields is just to implement the equations of motion in cohomology
(\cite{Fis_Hen_Sta_Tei,Hen_Tei}).

(iii) Antifield dependent solutions of $sA=0$ exist for simple gauge
groups but only in ghost degree $\geq 2$. For instance,
$\int j^\mu tr (A_\mu C^2)\ d^4x  + antifield\ dependent\ terms$ is such a
solution -which involves again a conserved current.

(iv)  Although of mathematical interest, it is not known whether the
solutions (\ref{currents}) and (\ref{anom_suppl}) do arise in practice
in the renormalization of integrated, gauge invariant operators.  An
example where an antifield-independent
 solution of the type $\mu_A C^A$ in (\ref{DRM}) occurs
is given in \cite{DixRM}.

(v) If there are free abelian gauge fields, the equation (\ref{fund_eq})
admits further solutions. This case is rather academical in the
present context but is of interest when one considers the problem
of introducing consistent couplings among free, spin 1 gauge fields
 \cite{Bar_Hen}. The extra solutions correspond to
non-abelian deformations of the abelian theory. They are not
written here for the sake of brievity but will be discussed elsewhere
\cite{Bar_Hen2}.

We now pass to the demonstration of the theorems. The BRST differential
in Yang-Mills theory is a sum of two differentials,
\begin{eqnarray}
s=\delta +\gamma\label{decomposition}
\end{eqnarray}
Since both $\delta$ and $\gamma$ are derivations and commute with
$\partial_\mu$, it is enough to define $\delta$ and $\gamma$ on the
generators $A^a_\mu$, $C^a$, $y^i$, $A^{*\mu}_a$, $C^*_a$ and
$y^*_i$ \cite{note}. One has
\begin{eqnarray}
\delta A^a_\mu=0,\ \delta C^a=0,\ \delta y^i=0\\
\delta A^{*\mu}_a={\delta{\cal L}\over\delta A^a_\mu},
\ \delta C^*_a=D_\mu A^{*\mu}_a - T^j_{ai}y^*_j y^i,
\ \delta y^*_i= {\delta{\cal L}\over\delta y^i}
\end{eqnarray}
and
\begin{eqnarray}
\gamma A^a_\mu=D_\mu C^a,\ \gamma C^a={1\over 2} C^a_{bc}C^b C^c,
\ \gamma y^i=T^i_{aj}y^jC^a\\
\gamma A^{*\mu}_a=-A^{*\mu}_c C^c_{ab}C^b,\ \gamma C^*_a=
-C^*_c C^c_{ab}C^b,\nonumber\\ \gamma y^*_i=-T^j_{ai}y^*_jC^a ,
\end{eqnarray}
where the $T_a$'s are the generators of the representation to
which the $y^i$ belong.  This implies
\begin{eqnarray}
\delta^2=0,\ \gamma^2=0,\ \gamma\delta+\delta\gamma=0 .
\end{eqnarray}
The decomposition (\ref{decomposition}) of $s$ is quite standard from the
point of view of the antifield formalism
\cite{Fis_Hen_Sta_Tei,Hen_Tei}. It
corresponds to the introduction of a further grading besides the
ghost number,which is called the ``antighost" (or ``antifield")
number and is denoted by $antigh$,
\begin{eqnarray}
antigh\ A^a_\mu=0,\ antigh\ C^a=0,\ antigh\ y^i=0\nonumber\\
antigh\ A^{*\mu}_a=1,\ antigh C^*_a
=2,\ antigh\ y^*_i=1\\ antigh\ \delta = -1,\ antigh\ \gamma =0\nonumber
\end{eqnarray}
(One has also, of course, $gh\ A^a_\mu=0$, $gh\ C^a=1$, $gh\ y^i=0$,
$gh\ A^{*\mu}_a=-1$, $gh C^*_a=-2$, $gh\ y^*_i=-1$,
$gh\ \delta = gh\ \gamma = gh\ s=1$). The differential
$s$ is sometimes called the full, unrestricted BRST differential,
while the differential $\gamma$ acting on the polynomials in $A^a_\mu,
C^a$ and their derivatives (but no antifields), is called the
restricted BRST differential \cite{Dix}. The differential $\delta$
is the ``Koszul-Tate differential" and provides a resolution of the
algebra of functions on the stationary surface where the equations of
motion hold \cite{Fis_Hen_Sta_Tei,Hen_Tei}.

The first step in the proofs of theorems 1 and 2 amounts to rewriting
the condition $s\int a\ d^4x =0$ in terms of the integrand $a$.
To that end, it is convenient to adopt form notations and to replace the
polynomial $a$ by the $4$-form $a\ dx^0\wedge\dots\wedge d^3x$, which
we still denote by $a$. We shall call ${\cal B}$ the algebra of
spacetime exterior forms with coefficients that are polynomials
in $A^a_\mu$, $C^a$, $y^i$, $A^{*\mu}_a$, $C^*_a$, $y^*_i$ and their
derivatives~; and we shall call ${\cal E}$ the algebra of spacetime exterior
forms with coefficients that are polynomials in $A^a_\mu$, $C^a$, $y^i$ and
their derivatives (no antifields).

Because of Stokes theorem ( $\int db =0$ -we assume as it is usual
in this context, that all surface terms are zero), one can remove
the integral sign at the price of allowing for the presence of a total
exterior derivative. The condition $s\int a =0$ is equivalent to
\begin{eqnarray}
s a+db=0\label{loc_fund_eq}.
\end{eqnarray}
And $\int a$ is exact if and only if
\begin{eqnarray}
a = sc +de .
\end{eqnarray}
Thus, the problem of computing the cohomology of $s$ in the space of
integrated polynomials becomes that of computing the cohomology of
$H^{*,4}(s|d,{\cal B})$ of $s$ modulo $d$ in the algebra of
polynomial forms of degree $4$. Note that the exterior derivative $d$
anticommutes with $\delta$ and $\gamma$ since $[\partial_\mu,\delta]=
[\partial_\mu,\gamma]=0$.

	A number of cohomologies related to $H^{*,4}(s|d,{\cal B})$ have
already been computed explicitely in the literature. These are
$H^*(d,{\cal B})$, $H^*(\delta,{\cal B})$, $H^*(\gamma,{\cal E})$,
$H^*(\gamma,{\cal B})$, $H^*(s,{\cal B})$,
$H^*(d,H^*(\gamma,{\cal E}))$ and $H^*(\gamma|d,{\cal E})$.
Of particular importance for the sequel is $H^*(\gamma|d,{\cal E})$
because any solution $a$ of $sa+db=0$ which does not involve the
antifields is automatically a solution of $\gamma a + db =0$.
And if it is trivial in $H^*(\gamma|d,{\cal E})$ (i.e., of the form
$\gamma c+de$ where $c$ and $e$ do not involve the antifields), then
it is also trivial in $H^*(s|d,{\cal B})$ ($\gamma c =sc$) \cite{note2}.
The strategy for computing $H^*(s|d,{\cal B})$ adopted here is to relate
as much as possible elements of $H^*(s|d,{\cal B})$ to the known elements
of $H^*(\gamma|d,{\cal E})$ \cite{Dix,Bra_Dra_Kre,Dub_Hen_Tal_Via} by
eliminating the antifields. This is not always possible. But the
equivalence classes of  $H^*(s|d,{\cal B})$ with no representative in
${\cal E}$ are easily exhibited and correspond precisely to
(\ref{currents}) and (\ref{anom_suppl}) above.

In order to control the antifield dependence of the solutions of
(\ref{loc_fund_eq}), one needs two intermediate  results.
\begin{lemma}
: let $b$ be a $\gamma$-closed form with degree $< 4$ and antighost
number $>0$. If $db$ is $\gamma$-exact, then $b$ is trivial,
i.e., $b=\gamma m + dn$ where $n$ is $\gamma$-closed. In other words,
\begin{eqnarray}
H^{l,j}_k(d,H^*(\gamma,{\cal B}))=0\ for\ j<4\ and\ k\geq 1.
\end{eqnarray}
\end{lemma}
Here and in the sequel, the first upper index $l$ is the ghost number, the
second index $j$ is the form degree and the lower index $k$ is the
antighost number. The lemma is proved exactly as the proposition
on page 363 of \cite{Dub_Hen_Tal_Via}.

The second intermediate result deals with $H^{l,j}_k(\delta|d)$.
It is well known that $H^{l,j}_k(\delta)$ vanishes for $k\geq 1$
in ${\cal B}$ \cite{Hen2}. It turns out that the cohomological groups
$H^{l,j}_k(\delta|d)$ with $k\geq 1$ are not zero, but only for $k=1$
and $j=4$. More precisely, one has
\begin{lemma}
: The group $H^{-1,4}_1(\delta|d)$ is isomorphic to the space of non
trivial conserved currents. The other groups $H^{l,j}_k(\delta|d)$
with $k\geq 1$ all vanish.
\end{lemma}
This lemma will be proved in \cite{Bar_Hen2}.

We can now demonstrate the theorems. Let $a$ be a $4$-form
solution of $sa+db=0$ with ghost number equal to $0$ or $1$.
Expand $a$ and $b$ according to the antighost number,
\begin{eqnarray}
a=a_0+\dots+a_k,\ \ b=b_0+\dots+b_k.
\end{eqnarray}
Assume $k\geq 1$. The equation $sa+db=0$ implies, at antighost number
$k$, $\gamma a_k +db_k =0$. By the standard descent argument,
this equation yields the chain of equations $\gamma b_k +db^\prime_k =0$,
$\gamma b^\prime_k +db^{\prime\prime}_k =0$, $\gamma b^{\prime\prime}_k
 +db^{\prime\prime\prime}_k =0$, $\gamma b^{\prime\prime\prime}_k =0$
for some $2$-form $b^\prime_k$, $1$-form $b^{\prime\prime}_k$ and
$0$-form $b^{\prime\prime\prime}_k$. By lemma 1, $b^{\prime\prime\prime}_k$
is trivial, $b^{\prime\prime\prime}_k=\gamma\rho_k$. But then one can make
redefinitions so that it vanishes.
Repeating the argument for  $b^{\prime\prime}_k$,
$b^\prime_k$ and $b_k$ shows that $b_k$ can be assumed to be equal to zero.
Thus, $a_k$ is $\gamma$-closed. The general solution of $\gamma a_k =0$
is known \cite{Dix,Bra_Dra_Kre,Dub_Hen_Tal_Via,Ban,Hen} and reads, up to
inessential $\gamma$-exact terms
\begin{eqnarray}
a_k=\Sigma a_J \omega^J(C)\label{poly_inv}
\end{eqnarray}
where $\omega^J(C)$ form a basis of the cohomology of the Lie algebra
of the gauge group (invariant cocycles) and where $a_J$ are invariant
polynomials in the field strengths, the matter fields, the antifields and
their covariant derivatives. The last but one equation in
$sa+db=0$ reads $\gamma a_{k-1} +\delta a_k +db_{k-1}=0$.
Since $\gamma\delta a_k=-\delta\gamma a_k =0$, one gets again
a descent for $b_{k-1}$, which reads
$\gamma b_{k-1} +db^\prime_{k-1} =0$,
$\gamma b^\prime_{k-1} +db^{\prime\prime}_{k-1} =0$,
$\gamma b^{\prime\prime}_{k-1} +db^{\prime\prime\prime}_{k-1} =0$,
$\gamma b^{\prime\prime\prime}_{k-1} =0$.
Two cases must be considered:

(i) $k-1\geq 1$. Repeated applications of lemma 1 show again that
$b^{\prime\prime\prime}_{k-1}$, $b^{\prime\prime}_{k-1}$ and
$b^\prime_{k-1}$ can be chosen to vanish so that $b_{k-1}$ is $\gamma$-closed,
\begin{eqnarray}
b_{k-1}=\Sigma b_J \omega^J(C)\label{poly_inv1}.
\end{eqnarray}
Inserting the respective forms (\ref{poly_inv}) and (\ref{poly_inv1}) of
$a_k$ and $b_{k-1}$ in $\gamma a_{k-1} +\delta a_k +db_{k-1}=0$ yields the
condition that $\Sigma (\delta a_J + d b_J) \omega^J(C)$ should be
$\gamma$-exact.
But then, $(\delta a_J + d b_J)$ must vanish. Lemma 2 implies that
$a_J$ is $\delta$-closed modulo $d$, $a_J=\delta p_J + dq_J$
(where $p_J$ and $q_J$ may be chosen to fulfill
$\gamma p_J=0=\gamma q_J$ \cite{Bar_Hen2}). Straightforward
redefinitions enable one to set $a_J$ -and thus $a_k$-
equal to zero. One can similarily set successively
$a_{k-1}, a_{k-2},\dots$ equal to zero, until one reaches $a_1$.

(ii) $k=1$. If the gauge group is semi-simple, then $a_1
=0$.  Indeed, there is no Lie algebra cohomology in degree
$1$ or $2$.  The equation $sa + db = 0$ reduces to
$\gamma a_0 + db_0 =0$, which is precisely
the equation studied and solved in
the literature \cite{Dix,Bra_Dra_Kre,Dub_Hen_Tal_Via}. The coresponding
solutions are (\ref{lagrangians}), (\ref{Adler-Bardeen}) and
(\ref{gammaa_mod_d}).
If the gauge group has abelian factors, there are
additional solutions because the Lie algebra cohomology is non
trivial in degree $1$ or $2$. The invariant cocycles are
$k^\Delta_A C^A$ in degree $1$ and
$k^\Delta_{AB} C^A C^B$ in degree $2$.  For each of these cocycles,
one may form an element $a_1$ by taking the product with an
element of $H^{-1, 4}_1 (\delta |d)$, which is isomorphic to the
space of non trivial conserved currents (lemma $2$).  The
corresponding solutions are (\ref{currents}) and (\ref{anom_suppl}).
 This completes the proof of the theorems.

In this letter, we have settled down some old problems on the
perturbative renormalization of the Yang-Mills field with a
simple or semi-simple gauge group.  We
have shown that the inclusion of the antifields (sources for
the BRST variations) does not modify the BRST cohomology in
the space of integrated polynomials with ghost
degree equal to $0$ or $1$, except for making
trivial BRST invariant objects that vanish on-shell. The
BRST cohomology in higher ghost degree is modified, however.
We have also analyzed the
case with abelian factors and have exhibited all BRST invariant
integrated polynomials.  These may now depend non trivially
on the antifields, as in \cite{Bra}.  The new solutions are of
mathematical interest in the sense that their understanding involve
many ingredients (conserved currents, Koszul-Tate resolution ...) but
we have not investigated whether they actually occur.
Finally, the same techniques can be
applied to the calculation of $H^{l,j}(s|d,{\cal B})$ for any ghost
number $l$ or form degree $j$ and in any number of spacetime dimensions.
A fuller account of our work will be reported elsewhere \cite{Bar_Hen2}.

Useful conversations with F. Brandt, M. Dubois-Violette, M. Talon and
C.M. Viallet are gratefully acknowledged. G.B. is Aspirant au Fonds National
de la Recherche Scientifique (Belgium). This work has been partly supported
by a FNRS research grant and by a research contract with the Commission of
the European Communities.


\begin{thebibliography}{999}
\bibitem{note} The differential $\gamma$ was denoted by $\sigma$
in \cite{Hen}. However, since the letter $\sigma$ is also
used for the ``quantum BRST differential" $\sigma = s + \hbar \Delta$
\cite{Hen_Tei}, we shall not use this notation here in order to
avoid confusion.
\bibitem{note2} The converse is not true. One may have $a\not= \gamma
c+de$ but $a=sc +de$, for instance if $a$ vanishes with the equations of
motion.
\end{thebibliography}
\end{document}